\documentclass[prd,onecolumn,nogroupedaddress, nofootinbib, notitlepage, 10pt]{revtex4}  
\usepackage{graphicx}
\usepackage{amsmath}
\usepackage{amsfonts}
\usepackage{amssymb}
\usepackage{appendix}
\usepackage{mathtools}
\usepackage{comment}
\usepackage{bbold}
\usepackage{color}
\usepackage{slashed}
\usepackage{hyperref}
\usepackage{subfigure}
\usepackage{setspace}
\usepackage{enumitem}
\usepackage{longtable}
\usepackage{wasysym}
\usepackage[usenames,dvipsnames]{xcolor}
\usepackage{bm}
\usepackage[letterpaper, margin=.8in, top=0.85in, bottom=0.85in]{geometry}

\begin{document}

\singlespacing

\title{Baryon Number, Lepton Number, and Operator Dimension in the Standard Model}
\author{Andrew Kobach}
\affiliation{Department of Physics, University of California, San Diego, La Jolla, CA 92093, USA}

\begin{abstract}
We prove that for a given operator in the Standard Model (SM) with baryon number $\Delta B$ and lepton number $\Delta L$, that the operator's dimension is even (odd) if $(\Delta B - \Delta L)/2$ is even (odd).  Consequently, this establishes the veracity of statements that were long observed or expected to be true, but not proven, e.g., operators with $\Delta B- \Delta L=0$ are of even dimension, $\Delta B - \Delta L$ must be an even number, etc.  These results remain true even if the SM is augmented by any number of right-handed neutrinos with $\Delta L=1$. 
\end{abstract}

\date{\today}

\maketitle


The Standard Model (SM) has been remarkably successful at providing accurate expectations for experimental results, but it remains agnostic regarding a number observations, e.g., neutrino masses, dark matter, baryon asymmetry, etc.  
The SM must be extended in order to account for these shortcomings, but not at the expense of undoing the predictive power provided by the $SU(3)_c \otimes SU(2)_L \otimes U(1)_Y$ group symmetries in the SM. 
An interesting peculiarity among operators in the renormalizable Lagrangian of the SM is that baryon number $\Delta B$ and lepton number $\Delta L$ appear as accidental global symmetries, which are not violated in perturbation theory.  
%
%
A phenomenological implication of the baryon asymmetry of the universe and nonzero neutrino masses may be that there are additional degrees of freedom that violate $\Delta B$ and $\Delta L$, beyond those already present in the SM.

To date, no compelling evidence exists for new particles with mass $\lesssim$ 1 TeV, and there are strong experimental constraints on the scale for new degrees of freedom that could mediate $\Delta B$ and $\Delta L$ processes, like nucleon decay and neutrinoless double beta decay. 
Given these facts, the simplest way to parameterize $\Delta B$- and $\Delta L$-violating processes is via effective field theory, where heavy degrees of freedom are integrated out, and the SM can be augmented with effective operators of mass dimension $d>4$.  
If so, one can parameterize processes that violate $\Delta B$ and $\Delta L$ with effective operators while still preserving the gauge symmetries of the SM.  
Such a parameterization ignores details of physics high energies and permits a democratic treatment of operators that can be used to extend the SM.

Some surveys of the effective operators in the SM notice that there are distinct patterns between the mass dimension $d$ of the effective operator and its value of $\Delta B$ and $\Delta L$~\cite{Weinberg:1979sa, Wilczek:1979hc, Weinberg:1980bf, Weldon:1980gi, Rao:1982gt, Rao:1983sd, Buchmuller:1985jz, Babu:2001ex, deGouvea:2007qla, Grzadkowski:2010es, Degrande:2012wf, Angel:2012ug, Alonso:2013hga, deGouvea:2014lva, Lehman:2014jma, Henning:2015alf, Lehman:2015coa}.  
The most prominent observation is that for a given value of $\Delta B$ and $\Delta L$, there is a minimum value of $d$.  
For example, given an operator with $|\Delta L|=2$ and $\Delta B=0$, the lowest value of $d$ (when there are no right-handed neutrinos) is 5, specifically, the Weinberg operator, $(LH)^2$, and consequently other operators with $|\Delta L|=2$, $\Delta B=0$ have odd dimension, since a tower of higher-dimensional operators can be constructed by adding on the dimension-2 SM operator $(H^*H)$.  However, it remains to be proven deductively that $|\Delta L|=2$, $\Delta B=0$ operators {\it cannot} have even mass dimension. 
Other well-known observations are that $d=6$ operators have $\Delta B-\Delta L=0$~\cite{Weinberg:1979sa,Wilczek:1979hc, Weinberg:1980bf, Buchmuller:1985jz, Grzadkowski:2010es, Alonso:2013hga, Henning:2015alf}, operators with odd dimension must violate $\Delta B$ or $\Delta L$~\cite{Rao:1983sd, Degrande:2012wf, Lehman:2014jma, Henning:2015alf}, etc.  
General statements like these about expectations regarding the dimension of the operators that give rise to processes that violate baryon and lepton number can greatly simplify the process of model building and effective operator analysis.

It may be useful to go beyond noticing patterns after writing down every operator in the SM for a fixed value of $d$, as done in Refs.~\cite{Weinberg:1980bf, Buchmuller:1985jz, Grzadkowski:2010es, Alonso:2013hga, Lehman:2014jma, Henning:2015alf, Lehman:2015coa}, and instead deduce what relationships between $d$, $\Delta B$, and $\Delta L$ must hold, given only the requirements of invariance under SM gauge symmetries and Lorentz transformations.    
There have been some attempts at deriving a relationship between $\Delta B$, $\Delta L$, and $d$~\cite{Rao:1983sd, Degrande:2012wf, deGouvea:2014lva}, but none so far have been sufficiently general in order to prove all possible statements about any operator dimension $d$ and any value of $\Delta B$ or $\Delta L$.  
Here, we provide a proof that for a given operator in the SM, that $d$ is even (odd) if $|\Delta B -  \Delta L|/2$ is even (odd), which follows directly from Lorentz invariance and $U(1)_Y$ hypercharge invariance.  These results remain valid even if there are any number of right-handed neutrinos with $\Delta L$=1 added to the particle content of the SM.


All operators in the SM are built out of matter fields ($Q, u^c, d^c, L, e^c, Q^\dagger, u^{c\dagger}, d^{c\dagger}, L^\dagger, e^{c\dagger}$),\footnote{Of course, there are three generations of matter fields, though because anomaly cancelation is required for each generation independently, we can consider here only a single family without loss of generality.  Also, it is convenient for our purposes, and somewhat conventional, to leave implicit the identifications of the color charges, i.e., red, blue, green, for the quark fields.  As long as there is the correct {\it number} of quark fields, then we presume that a specific assignment of color charges can be picked to form an $SU(3)_c$ singlet.  This is a common simplification, because only the number of colors, and not specific colors themselves, is physically observable. } covariant derivatives $D_\mu$, Higgs fields ($H$, $H^*$), and field strength tensors ($B_{\mu\nu}, W_{\mu\nu}, G_{\mu\nu}$).  
In addition to the known SM degrees of freedom, there may exist another chiral neutrino: $\nu^c$ (along with its complex conjugate, $\nu^{c\dagger}$), which is often utilized, though it is not absolutely necessary, to explain the observation that neutrinos have nonzero mass.    
When these building blocks are put together to form an operator of any mass dimension in the SM, one must conserve Lorentz invariance and ensure that the operator transforms trivially under the $SU(3)_c \otimes SU(2)_L \otimes U(1)_Y$ group symmetries.  

These requirements for an SM operator can be easily imposed by noting that they can be interpreted as global constraints on the number $N$ of fermion fields, Higgs fields, covariant derivatives, and field-strength tensors in the operator.  
For example, the requirement that the operator is invariant under the $U(1)_Y$ weak hypercharge symmetry (hypercharge henceforth) can be stated as\footnote{The hypercharge assignments in Eq.~(\ref{hypercharge}) are twice the value of the assignments typically seen in the literature.  This is purely convention, and the hypercharge normalization used in this analysis is chosen just for the purposes of algebraic simplification. }
\begin{equation}
\label{hypercharge}
0 = \frac{1}{3} \left( N_Q - N_{Q^\dagger} \right) - \frac{4}{3} \left( N_u - N_{u^\dagger} \right) + \frac{2}{3} \left( N_d - N_{d^\dagger} \right) - \left( N_L - N_{L^\dagger} \right) + 2\left( N_e - N_{e^\dagger} \right) + \left( N_H - N_{H^*} \right)
\end{equation}
Interestingly, the constraints placed on an SM operator by requiring $SU(3)_c \otimes SU(2)_L$ invariance are, in fact, contained in Eq.~(\ref{hypercharge}). To see how, note that because all $N$'s must be a positive integer, and the last three terms in Eq.~(\ref{hypercharge}) sum to an integer, it immediately follows that the sum of the first three terms in Eq.~(\ref{hypercharge}) must also be an integer, which can be rewritten as
\begin{equation}
\frac{1}{3}\left(N_Q + N_{d^\dagger} + N_{u^\dagger} \right) - \frac{1}{3}\left(N_{Q^\dagger} + N_{d} + N_{u} \right) - (N_u - N_{u^\dagger}) + (N_d - N_{d^\dagger}) \in \mathbb{Z},
\end{equation}
and it immediately follows that
\begin{equation}
\label{su3constraint}
 \frac{1}{3}\left(N_Q + N_{d^\dagger} + N_{u^\dagger} \right) - \frac{1}{3}\left(N_{Q^\dagger} + N_{d} + N_{u} \right) \in \mathbb{Z}. 
 \end{equation}
Eq.~(\ref{su3constraint}) is exactly the requirement that the effective operator is invariant under $SU(3)_c$ transformations.    
Likewise, we can begin again with Eq.~(\ref{hypercharge}), and rewrite it as
\begin{equation}
\label{hyper2}
0 = \left( N_Q - N_{Q^\dagger} \right) - 3\left( N_L - N_{L^\dagger} \right) + 3\left( N_H - N_{H^*} \right) - 4 \left( N_u - N_{u^\dagger} \right) + 2 \left( N_d - N_{d^\dagger} \right)  + 6\left( N_e - N_{e^\dagger} \right) .
\end{equation}
Since the last three terms in Eq.~(\ref{hyper2}) must sum to an even integer, the sum of the first three terms must also sum to an even integer.  
To show this will always hold, here we can note that the requirement of invariance under the $SU(2)_L$ transformations is no more than the requirement that there are an even number of $SU(2)_L$ doublets:  
\begin{equation}
\label{SU2constraint}
\text{even} = N_Q + N_{Q^\dagger}  + N_L + N_{L^\dagger} + N_H + N_{H^*}.
\end{equation}
Since an even number of fields can be added to both sides of Eq.~(\ref{SU2constraint}), it implies that
\begin{equation}
\text{even} = \left(N_Q - N_{Q^\dagger}\right)  -3\left(N_L - N_{L^\dagger}\right) + 3\left(N_H - N_{H^*}\right)
\end{equation}
should also hold.  Therefore, Eq.~(\ref{hypercharge}) also contains the $SU(2)_L$ constraint.  This shows that the hypercharge assignments for the fields in the SM contain all the necessary information needed when determining whether a given operator is a SM singlet under $SU(3)_c \otimes SU(2)_L \otimes U(1)_Y$ group transformations.  Thus, we only make use of only the $U(1)_Y$ hypercharge constraint to ensure gauge invariance.

We now turn to the discussion of Lorentz invariance.  If there are no $\sigma^\mu$ matrices in the operator, then the number of right- and left-handed fields must each be even in order for the operator to be Lorentz invariant.  Because there must be at least one right- and left-handed field each for every $\sigma^\mu$ matrix in the operator to form a Lorentz singlet, it implies that if there are an even (odd) number of $\sigma^\mu$ matrices in the operator, then there are an even (odd) number of right-handed fields {\it and} an even (odd) number of left-handed fields.  
Additionally, we can note that the only way to contract a Lorentz index associated with a derivative is to contract it with either a $\sigma^\mu$ matrix or a field strength tensor.  Because a $\sigma^\mu$ matrix has one Lorentz index and a field strength tensor has two, an even (odd) number of Lorentz indices associated with derivatives must correspond to an even (odd) number of $\sigma^\mu$ matrices in the operator.  If the operator has an odd number of derivatives, then the operator must have an odd number of $\sigma^\mu$ matrices.  
Therefore, from these observations, Lorentz invariance can be stated as 
\begin{equation}
\label{lorentzLH}
\text{if } N_D \text{ is even (odd), then } (N_{Q} + N_{u} + N_{d}  + N_{L} + N_{e} + N_{\nu}) \text{ is even (odd)} ,
\end{equation}
and
\begin{equation}
\label{lorentzRH}
\text{if } N_D \text{ is even (odd), then } (N_{Q^\dagger} + N_{u^\dagger} + N_{d^\dagger}  + N_{L^\dagger} + N_{e^\dagger} + N_{\nu^\dagger}) \text{ is even (odd)} .
\end{equation}
Here, $N_D$ stands for the number of covariant derivatives, and $N_\nu$ ($N_{\nu^\dagger}$) counts the total number of left-handed antineutrinos (right-handed neutrinos).  Together, the constraint of hypercharge invariance, Eq.~(\ref{hypercharge}), and Lorentz invariance, Eqs.~(\ref{lorentzLH}) and (\ref{lorentzRH}), constitute the necessary and sufficient conditions for operators in the SM.

In order to discuss any relationships between $\Delta B$, $\Delta L$, and the operator mass dimension $d$, we begin by defining $d$ as 
\begin{eqnarray}
\label{dim}
d = \frac{3}{2} \left( N_Q + N_{Q^\dagger} + N_u + N_{u^\dagger} + N_d + N_{d^\dagger} + N_L + N_{L^\dagger} + N_e + N_{e^\dagger} + N_\nu + N_{\nu^\dagger} \right) + N_H + N_{H^*}  + N_D + 2N_X, 
\end{eqnarray}
Here,  $N_X$ indicates the number of field-strength tensors, i.e., $B_{\mu\nu}$, $W_{\mu\nu}$, or $G_{\mu\nu}$.  Given the definitions of baryon number $\Delta B$ and lepton number $\Delta L$,
\begin{eqnarray}
\label{Ldef}
 \Delta L &\equiv& N_L + N_{e^\dagger} + N_{\nu^\dagger} - \left(N_{L^\dagger} + N_e + N_\nu \right) , \\
 \label{Bdef}
 \Delta B &\equiv& \frac{1}{3} \left( N_Q + N_{u^\dagger} + N_{d^\dagger}\right) - \frac{1}{3} \left( N_{Q^\dagger} + N_{u} + N_{d}\right),
\end{eqnarray}
we know from its definition that $\Delta L$ is an integer ($\Delta L\in \mathbb{Z}$), and from Eq.~(\ref{su3constraint}) that hypercharge invariance implies that $\Delta B$ must be an integer ($\Delta B \in \mathbb{Z}$).  
From the definition of $d$ in Eq.~(\ref{dim}) and the fact that $\Delta B \in \mathbb{Z}$, we can note that for an operator with a given value of $\Delta B$ and $\Delta L$, the minimum value of its mass dimension $d_\text{min}$ is
\begin{equation}
\label{dmin}
d_\text{min} \geq \frac{9}{2} |\Delta B| + \frac{3}{2} |\Delta L| .
\end{equation}
Eq.~(\ref{dmin}) is a weak lower bound, but is an exact equality when the operator contains only fermionic matter fields and when $\Delta B$ or $\Delta L$ are nonzero. While it is commonly noted that there is a value of $d_\text{min}$ for a given value of $\Delta B$ and $\Delta L$, e.g., when $\Delta B = 0$ and $|\Delta L| = 2$, $d_\text{min} = 3~ (5)$ if there are (no) right-handed neutrinos~\cite{Weinberg:1979sa}, it is not trivial to determine whether or not there is an analytical function for $d_\text{min}$ as a function of only $\Delta B$ and $\Delta L$.  Finding such a solution would require solving a discrete optimization problem, but with 16 free parameters subject to a few constraints, i.e., hypercharge invariance, Lorentz invariance, and definitions of $\Delta B$ and $\Delta L$, it remains an open question whether there exists an analytical solution.

Since one may not be able to derive a closed analytical expression for $d_\text{min}$, the next best thing may be to determine whether $d$ is even or odd, given only values of $\Delta B$ and $\Delta L$.  
Inserting the constraint of hypercharge invariance in Eq.~(\ref{hypercharge}) and the definitions of $\Delta B$ and $\Delta L$ in Eqs.~(\ref{Bdef}) and (\ref{Ldef}) into the definition of $d$ in Eq.~(\ref{dim}) gives:
\begin{equation}
d = 3\left( N_{Q^\dagger} + 2N_u - N_{u^\dagger} + N_{d^\dagger} + N_{L^\dagger} + N_{e^\dagger} + N_{\nu^\dagger}  + \frac{1}{2}  \Delta B + \frac{3}{2} \Delta L \right) - 2N_H + 4 N_{H^*} + N_D + 2N_X .
\end{equation}
At this point, determining whether $d$ is even or odd requires determining whether the number 
\begin{equation}
\label{evenorodd}
3\left(N_{Q^\dagger} - N_{u^\dagger} + N_{d^\dagger} + N_{L^\dagger} + N_{e^\dagger} + N_{\nu^\dagger}  \right) + N_D ,
\end{equation}
is even or odd.  
The requirement of Lorentz invariance in Eq.~(\ref{lorentzRH}) proves, in fact, that Eq.~(\ref{evenorodd}) is always even.  Because $d$ is a positive integer ($d\in \mathbb{N}$), one therefore can conclude:
\begin{equation}
\label{result1}
\left( \frac{1}{2} \Delta B + \frac{3}{2}  \Delta L \right) \in \mathbb{Z}~  \bigg\{ \begin{array}{ll} \text{even}, & d\rightarrow \text{even} \\ \text{odd}, & d\rightarrow \text{odd} \end{array}
\end{equation}
Or, equivalently,
\begin{equation}
\label{result2}
\frac{\left( \Delta B -  \Delta L \right)}{2} \in \mathbb{Z}~  \bigg\{ \begin{array}{ll} \text{even}, & d\rightarrow \text{even} \\ \text{odd}, & d\rightarrow \text{odd} \end{array}
\end{equation}
These statements can be further summarized succinctly by the following statements:
\begin{eqnarray}
\label{evennums}
\text{$d$ is even} &\longleftrightarrow& |\Delta B - \Delta L| = 0, 4, 8, 12, ... \\
\label{oddnums}
\text{$d$ is odd} &\longleftrightarrow& |\Delta B - \Delta L| =  2, 6, 10, 14,...
\end{eqnarray}
There are some useful takeaways, that are direct consequences from Eq.~(\ref{dmin}) and Eqs.~(\ref{result1}) - (\ref{oddnums}):
\begin{itemize}
\item $|L|=2$ operators responsible for Majorana neutrino masses occur at odd mass dimension.  This was observed to be true in Refs.~\cite{deGouvea:2007qla, Babu:2001ex, Angel:2012ug} for those operators up to and including $d=11$, but not including those that contain derivatives, gauge bosons, nor non-trivial Lorentz structure.  
\item If $\Delta B - \Delta L = 0$, the operator must be of even mass dimension.  Specifically, operators with $d=6$ and $d=8$ must have $\Delta B-\Delta L=0$.  This is validated for all operators with $d=6$ in Refs.~\cite{Buchmuller:1985jz, Grzadkowski:2010es, Alonso:2013hga, Henning:2015alf} and $d=8$ in Refs.~\cite{Lehman:2015coa, Henning:2015alf}.
\item Operators in the SM with odd mass dimension have either nonzero $\Delta B$ or $\Delta L$.  This was proven in Ref.~\cite{Degrande:2012wf} and observed to be true when $d=7$ in Refs.~\cite{Lehman:2014jma, Henning:2015alf} for the SM without right-handed neutrinos. 
\item If an operator comprised of only $N_f$ number of fermion fields, where $N_f/2$ is odd, then it has nonzero $(\Delta B - \Delta L)$.  This was proven in Ref.~\cite{Rao:1983sd}. 
\item Nucleon decays where $\Delta B=-\Delta L $ can be described by operators with odd dimension.  This was verified for $d=7$~\cite{Weinberg:1980bf, Weldon:1980gi}. 
\item If $|\Delta B|=2$, $\Delta L=0$, which is the case for $n-\bar{n}$ oscillations, then $d$ must be odd, and, specifically, $d_\text{min}=9$~\cite{Rao:1982gt, Rao:1983sd}.  
\item An operator with even (odd) $\Delta B$ must also have even (odd) $\Delta L$. 
\item Neither the value of $(\Delta B - \Delta L)$ nor $(\Delta B + \Delta L)$ can be an odd number for any operator in the SM. 
\end{itemize}
To reiterate, these results are a direct consequence of only two assumptions:~hypercharge invariance and Lorentz invariance, and remain true if any number of right-handed neutrinos augments the SM.  Similar results were found for the small subset of SM operators that contain no covariant derivatives, no field strength tensors, and have trivial Lorentz structure, i.e., no $\sigma^\mu$ matrices~\cite{deGouvea:2014lva}.

These results only apply to extensions of the SM that utilize the SM particle content within operators invariant under the SM gauge symmetries, i.e., effective operators.  Of course, there many instances of model building beyond the SM that do not fall into this category, e.g., if the SM is expanded in a way that (1) introduces new particles that are charged under the SM gauge symmetries, (2) the SM gauge symmetries are embedded within a larger group, (3) contain particles with non-trivial $\Delta B$ or $\Delta L$ assignments, and the list goes on.  
These types of possibilities are necessary to consider if one wishes to build a renormalizable model that can give rise to $\Delta B$- or $\Delta L$-violating processes.

\begin{acknowledgements}
AK is grateful for useful conversations and feedback from Andr\'{e} de Gouv\^{e}a, Ben Grinstein, and Aneesh Manohar.  This work is funded in part by DOE grant \#de-sc0009919.  
\end{acknowledgements}

\bibliography{bib}{}

\end{document}